\documentclass[aps,prl,reprint,twocolumn,showpacs,superscriptaddress]{revtex4-1}  
 
\usepackage{relsize}
\usepackage{amsmath}
\usepackage{amssymb}
\usepackage{epsfig}
\usepackage{graphicx}
\usepackage{hyperref}
\usepackage{dcolumn}   
\usepackage{tabu}
\usepackage{boldline}
\usepackage{slashed}
\usepackage{multirow}
\usepackage{color}
\usepackage[normal]{subfigure}
\usepackage{rotating}
\usepackage[margin=0.9in,a4paper]{geometry}
\usepackage[table]{xcolor}
\usepackage{enumitem}
\usepackage[utf8x]{inputenc}
\usepackage{colortbl}
\usepackage{array,multirow}
\definecolor{nicered}{rgb}{0.7,0.1,0.1}
\definecolor{nicegreen}{rgb}{0.1,0.5,0.1}
\definecolor{red}{rgb}{1.0, 0, 0}
\def\mX{{\mathcal{X}}}

\def\eq#1{{Eq.~(\ref{#1})}}

\def\eqssmall#1#2{{eqs.~(\ref{#1})--(\ref{#2})}}
\def\fig#1{{Fig.~\ref{#1}}}

\def\Table#1{{Table~\ref{#1}}}

\def\abs#1{\left| #1\right|}

\def\Tr{\mbox{Tr}\,}

\def\zb{\mathbb{Z}}
\def\zN{\mathbb{Z}_\mathbb{N}}
\def\gsim{\raise0.3ex\hbox{$\;>$\kern-0.75em\raise-1.1ex\hbox{$\sim\;$}}}
\def\lsim{\raise0.3ex\hbox{$\;<$\kern-0.75em\raise-1.1ex\hbox{$\sim\;$}}}

\def\mb[#1]{\mathbf{#1}}

\renewcommand{\bar}{\overline}

\definecolor{LightCyan}{rgb}{0.88,1,1}
\definecolor{piggypink}{rgb}{0.99, 0.87, 0.9}
\definecolor{applegreen}{rgb}{0.55, 0.71, 0.0}
\definecolor{darkpastelgreen}{rgb}{0.01, 0.75, 0.24}
\definecolor{green-yellow}{rgb}{0.68, 1.0, 0.18}

\newcommand{\beq}{\begin{equation}}
\newcommand{\eeq}{\end{equation}}
\newcommand{\beqa}{\begin{eqnarray}}
\newcommand{\eeqa}{\end{eqnarray}}
\newcommand{\Sec}[1]{ \medskip \noindent {\sl \bfseries #1}}
\newcommand{\subsec}[1]{ \medskip \noindent {\sl \bfseries #1}}

\newcommand{\eqn}[1]{eq.~(\ref{#1})}

\begin{document}


\title{Redefining the Axion Window}

\author{Luca Di Luzio}
\email{luca.di-luzio@durham.ac.uk}
\affiliation{\normalsize \it 
Institute for Particle Physics Phenomenology, Department of Physics, Durham University, DH1 3LE, United Kingdom}
\author{Federico Mescia}
\email{mescia@ub.edu}
\affiliation{\normalsize\it Dept.~de F\'{\i}sica Qu\`antica i
  Astrof\'{\i}sica, Institut de Ci\`encies del Cosmos (ICCUB), 
Universitat de Barcelona, Mart\'i Franqu\`es 1, E08028 Barcelona, Spain
}
\author{Enrico Nardi}
\email{enrico.nardi@lnf.infn.it}
\affiliation{\normalsize\it INFN, Laboratori Nazionali di Frascati, C.P.~13, 100044 Frascati, Italy}

\begin{abstract}
  \noindent
  A major goal of axion searches is to reach inside the parameter
  space region of realistic axion models. Currently, the boundaries of
  this region depend on somewhat arbitrary criteria, and it would be
  desirable to specify them in terms of precise phenomenological
  requirements.  We consider hadronic axion models and classify the
  representations $R_Q$ of the new heavy quarks $Q$. By requiring that
  $i)$ the $Q$ are sufficiently short lived to avoid issues with long
  lived strongly interacting relics, $ii)$ no Landau poles are induced
  below the Planck scale, fifteen cases are selected, which define a
  phenomenologically preferred axion window bounded by a maximum
  (minimum) value of the axion-photon coupling about twice (four
  times) larger than commonly assumed.  Allowing for more than one
  $R_Q$, larger couplings, as well as complete axion-photon
  decoupling, become possible.
\end{abstract}


 \maketitle

 \Sec{Introduction.}  In spite of its indisputable success, the
 standard model (SM) is not completely satisfactory: it does not
 explain unquestionable experimental facts like dark matter (DM),
 neutrino masses, and the cosmological baryon asymmetry, and it
 contains fundamental parameters with highly unnatural values, like
 the the Higgs potential term $\mu^2$, the first generation Yukawa
 couplings $h_{e,u,d}$, and the strong CP violating angle
 $\abs{\theta}< 10^{-10}$.  This last quantity 
 is somewhat special: its value is stable with respect to higher order
 corrections \cite{Ellis:1978hq} (unlike $\mu^2$) and (unlike
 $h_{e,u,d}$~\cite{Hall:2014dfa}) it evades explanations based on
 environmental selection~\cite{Ubaldi:2008nf}.  Thus, seeking
 explanations for the smallness of $\theta$ independently of other
 ``small values'' problems is theoretically motivated. Basically, only
 three types of solutions exist.  The simplest possibility, a massless
 up-quark, is now ruled out~\cite{Aoki:2016frl,Olive:2016xmw}.  The
 so-called Nelson-Barr type of models~\cite{Nelson:1983zb,Barr:1984qx}
 either require a high degree of fine tuning, often comparable to
 setting $\abs{\theta} \lsim 10^{-10}$ by hand, or rather elaborated
 theoretical structures~\cite{Dine:2015jga}.  The Peccei-Quinn (PQ)
 solution~\cite{Peccei:1977ur,Peccei:1977hh,Weinberg:1977ma,Wilczek:1977pj},
 although it is not  
 completely free from
 issues~\cite{Kamionkowski:1992mf,Holman:1992us,Barr:1992qq}, arguably
 stands on better theoretical grounds.

Setting aside theoretical considerations, the question whether the PQ solution
is the correct one could be set experimentally by detecting the axion.
In order to focus axion searches, it is then very important to 
identify as well as possible 
the region of parameter space 
where realistic axion models live. 
The vast majority of search techniques are sensitive to the
axion-photon coupling $g_{a\gamma\gamma}$, which is inversely
proportional to the axion decay constant $f_a$.  Since the axion mass
$m_a$ has the same dependence, theoretical predictions and 
experimental exclusion limits
can be conveniently
presented in the $m_a$-$g_{a\gamma\gamma}$ plane.  The commonly
adopted {\em axion band} 
corresponds roughly to
$g_{a\gamma\gamma} \sim m_a \alpha/(2\pi f_\pi m_\pi) \sim 10^{-10}\,
(m_a/{\rm eV})\,$GeV$^{-1}$
with a somewhat arbitrary width, chosen to include representative
models~\cite{Kaplan:1985dv,Cheng:1995fd,Kim:1998va}.  In this Letter
we put forth a definition of a {\it phenomenologically preferred}
axion window as the region encompassing {\it hadronic} axion models
which $i)$ do not contain cosmologically dangerous
relics; $ii)$ do not induce Landau poles (LP)
below some scale $\Lambda_{LP}$ close to the Planck mass
$m_P = 1.2 \cdot 10^{19}$ GeV.  While all the cases we consider belong
to the KSVZ type of models~\cite{Kim:1979if,Shifman:1979if}, the
resulting window encompasses also the DFSZ
axion~\cite{Zhitnitsky:1980tq,Dine:1981rt} and many of its
variants~\cite{Cheng:1995fd}.


\Sec{Hadronic axion models.} 
The basic ingredient of any renormalizable axion model is a global 
$U(1)_{PQ}$ symmetry. The associated Noether current $J_\mu^{PQ}$ must have 
a color anomaly and, although not required for solving the strong CP
problem,  in general it also has an electromagnetic anomaly:
\begin{equation}
\label{eq:NE}
\partial^\mu J_\mu^{PQ} = 
\frac{N \alpha_s}{4\pi} G^a_{\mu\nu} \tilde G^{a\mu\nu} + 
\frac{E \alpha}{4\pi} F_{\mu\nu} \tilde F^{\mu\nu} \,,
\end{equation}
where $G^a_{\mu\nu}\,(F_{\mu\nu})$ is the color (electromagnetic) field strength
tensor, $\tilde G^{a\mu\nu}\,(\tilde F^{\mu\nu}) = \tfrac{1}{2}
\epsilon^{\mu\nu\rho\sigma} G^a_{\rho\sigma}\,(F_{\rho\sigma})$ its dual, 
$N$ and $E$
the respective anomaly coefficients.  
In a generic axion model of KSVZ type~\cite{Kim:1979if,Shifman:1979if} the anomaly
is induced by pairs of 
heavy fermions $Q_L,\,Q_R$ which must
transform non-trivially under $SU(3)_C$ and chirally under $U(1)_{PQ}$.
Their mass arises from  a Yukawa interaction with a SM singlet
scalar $\Phi$ which develops a 
PQ breaking vacuum expectation value. Thus    
their PQ charges $\mX_{L,R}$,  
normalized to $\mX(\Phi)=1$, must satisfy
$|\mX_L-\mX_R|=1$.  
We denote the (vectorlike) representations of the SM gauge group
$G_{SM}\!\!=\!SU(3)_C\!\times\! SU(2)_I\!\times\! U(1)_Y$ to which we
assign the $Q$ 
as  $R_Q\!\!=\! (\mathcal{C}_Q,\mathcal{I}_Q,\mathcal{Y}_Q)$ 
so that
\beqa 
\label{N}
N & =&  \sum\nolimits_Q
 \left(\mX_L-\mX_R\right)
\,T(\mathcal{C}_Q) \,, \\   
 \label{E}
E &=&  \sum\nolimits_Q 
\left(\mX_L-\mX_R\right)\,\mathcal{Q}^2_Q \,,   
\eeqa 
where the sum is over irreducible color representations
(for generality we allow for the simultaneous presence of more
$R_Q$). The color index is defined by
$\Tr T_Q^a T_Q^b = T(\mathcal{C}_Q) \delta^{ab}$ with $T_Q$ the
generators in $\mathcal{C}_Q$ 
and $\mathcal{Q}_Q$ is the $U(1)_{\rm{em}}$ charge.  
The scalar field $\Phi$ can be parametrized as
\begin{equation}
\label{eq:Phi}
\Phi(x) = 
(1/\sqrt{2}) \left[\rho(x) + V_a\right] e^{i
  a(x)/V_a}\,. 
\end{equation} 
The mass of $\rho(x)$ is of  order  
$V_a \gg (\sqrt{2} G_F)^{-1/2} = 247\,$GeV, while a tiny mass for the axion $a(x)$
arises from nonperturbative QCD effects which explicitly
break $U(1)_{PQ}$.  The SM quarks $q=q_L,d_R,u_R$ do not contribute to
the QCD anomaly, and thus their PQ charges can be set to
zero.  
The renormalizable Lagrangian for a generic hadronic axion model can
be written as:
\begin{equation}
\label{LKSVZ}
\mathcal{L}_a = 
\mathcal{L}_{\rm SM}+
\mathcal{L}_{\rm PQ} 
 - V_{H \Phi} +
\mathcal{L}_{Qq} 
\,,
\end{equation}
where $\mathcal{L}_{\rm SM}$  is the SM Lagrangian, 
\begin{equation}
\label{LKSVZPQ}
\mathcal{L}_{\rm PQ} = |\partial_\mu \Phi|^2 + 
\overline{Q} i \slashed{D} Q 
- (y_Q \, \overline{Q}_L Q_R \Phi + \text{H.c.}) \, ,
\end{equation}
with $Q=Q_L+Q_R$.
The new scalar terms are:
\begin{equation}
\label{LKSVZV}
V_{H \Phi} 
= 
- \mu^2_\Phi  |\Phi|^2   + \lambda_\Phi |\Phi|^4 +  \lambda_{H\Phi}
|H|^2 |\Phi|^2 \,.
\end{equation} 
Finally, $\mathcal{L}_{Qq}$ contains possible renormalizable terms
coupling $Q_{L,R}$ to SM quarks 
 which can allow for $Q$  decays~\cite{Ringwald:2015dsf}. 
%
Note, however, that SM gauge invariance allows for 
$\mathcal{L}_{Qq} \neq 0$ only for a few specific $R_Q$.

\subsec{PQ quality and heavy $Q$ stability.}\  
The issue whether the $Q$ are exactly stable, metastable, or decay
with safely short lifetimes, is of central importance in our study,
so  let us discuss it in some detail. The gauge invariant kinetic term in
$\mathcal{L}_{\rm PQ}$ features a
$U(1)^3 \equiv U(1)_{Q_L} \times U(1)_{Q_R} \times U(1)_{\Phi}$
symmetry corresponding to independent rephasings of the $Q_{L,R}$ and
$\Phi$ fields.  The PQ Yukawa term ($y_Q\neq 0$) breaks
$U(1)^3$ to $U(1)^2$. One factor is the anomalous $U(1)_{PQ}$, 
the other one is a non-anomalous $U(1)_{Q}$,
that is the $Q$-baryon number of the new quarks~\cite{Kim:1979if}, 
 under which  $Q_{L,R} \rightarrow e^{i \beta} Q_{L,R}$ and  
$\Phi \rightarrow \Phi$. 
If $U(1)_Q$  were an exact symmetry, the new quarks 
would be absolutely stable. 
For the few
$R_Q$  for which $\mathcal{L}_{Qq}\neq 0$ is allowed,
$U(1)_Q\times U(1)_B$ is further broken to $U(1)_{B'}$, a generalized
baryon number extended to the $Q$, which can then decay with
unsuppressed rates.  However, whether $\mathcal{L}_{Qq}$ is allowed at
the renormalizable level, does not depend solely on $R_Q$, but also
on the specific PQ charges. For example,
independently of $R_Q$, the common assignment
$\mX_L=-\mX_R=\frac{1}{2}$ would forbid PQ invariant decay
operators at all orders. $U(1)_Q$ violating decays could then occur 
only via PQ-violating effective operators of dimension $d>4$.  
Both $U(1)_{PQ}$ and $U(1)_Q$ are expected to be  broken at least by
Planck-scale effects,  
inducing PQ violating contributions
to the axion potential $V^{d>4}_{\Phi}$ as well as an effective
Lagrangian $\mathcal{L}^{d>4}_{Qq}$. In particular, 
in order  
to preserve $\abs{\theta}<10^{-10}$, operators in $V^{d>4}_{\Phi}$ must be of
dimension
$d\geq 11$~\cite{Kamionkowski:1992mf,Holman:1992us,Barr:1992qq}.
Clearly, if $\mathcal{L}^{d>4}_{Qq}$ had to respect $U(1)_{Q}$  
to a similar level of accuracy, the $Q$'s would behave as effectively
stable.  However, a scenario in which 
$U(1)_Q$
arises as an accident because of specific assignments for the charges
of another global symmetry $U(1)_{PQ}$, seems theoretically untenable.
A simple way out is to assume a suitable discrete (gauge) symmetry
$\zN$ ensuring  that $i)$ $U(1)_{PQ}$ arises accidentally and
is of the required {\it high quality}; $ii)$ $U(1)_Q$ is either broken at
the renormalizable level, or it can be of sufficient {\it bad quality} to
allow for safely fast $Q$ decays.

\Table{zncharges} gives a neat example of how such a 
mechanism can work (see also \cite{Ringwald:2015dsf}). We choose $R_Q= R_{d_R}=(3,1,-1/3)$ so that 
$G_{SM}$ invariance allows for  $\mathcal{L}_{Qq}\neq 0$, 
and we assume the following transformations under $\zN$:
 $ Q_L \rightarrow Q_L \, , \quad Q_R \rightarrow \omega^{\mathbb{N}-1} Q_R \, ,
  \quad \Phi \rightarrow \omega\, \Phi \, ,$
%
%
with $\omega \equiv e^{i 2 \pi/\mathbb{N}}$.  
This ensures that the minimum dimension of the PQ breaking
operators in $V^{d>4}_{\Phi}$ is  $\mathbb{N}$.  
The dimension of $U(1)_Q$ breaking decay operators
depends on the $\zN$ charges of the SM quarks.  \Table{zncharges}
lists different possibilities for $d\leq 4$ and $d=5$.
The last column gives the PQ charges that one has to assign to $Q_{L,R}$ 
so that 
$U(1)_{PQ}$ 
can be defined also in the presence 
of the operators in column 2 and 3. 
\begin{table}[t!] 
 \renewcommand{\arraystretch}{1.0}
\centering
\begin{tabular}{@{} |c|c|c|c| @{}}
\hline
$\zN(q)$ &  $d\leq 4$ & $d=5$ & $(\mX_L,\mX_R)$  \\ 
\hline
\hline
$1$
& $\overline{Q}_L  d_R$ & 
$\overline{Q}_L \gamma_\mu q_L\left(D^\mu H\right)^\dag$
& $(0,-1)$  \\
\hline
$\omega$
& $\overline{Q}_L d_R \Phi^\dag$ &
& $(-1,-2)$  \\
\hline
$\omega^{\mathbb{N}-2}$ & --  &  
$\overline{Q}_L d_R\Phi^2$,\,
$\overline{Q}_R q_L H^\dag \Phi$
& $(2,1)$  \\
\hline
$\omega^{\mathbb{N}-1}$ & $\overline{q}_L Q_R H,\,  \overline{Q}_L d_R \Phi$ & 
-- 
& $(1,0)$  \\
\hline
  \end{tabular}
  \caption{\label{zncharges} 
    $\zN$ charges for the SM quarks $q$ which allow for 
    $d\leq 4$  and $d=5$  operators for   $R_Q=(3,1,-1/3)$.
  }
\end{table}

\begin{figure}[t!]
\centering
\includegraphics[angle=0,width=7.5cm,height=6.8cm]{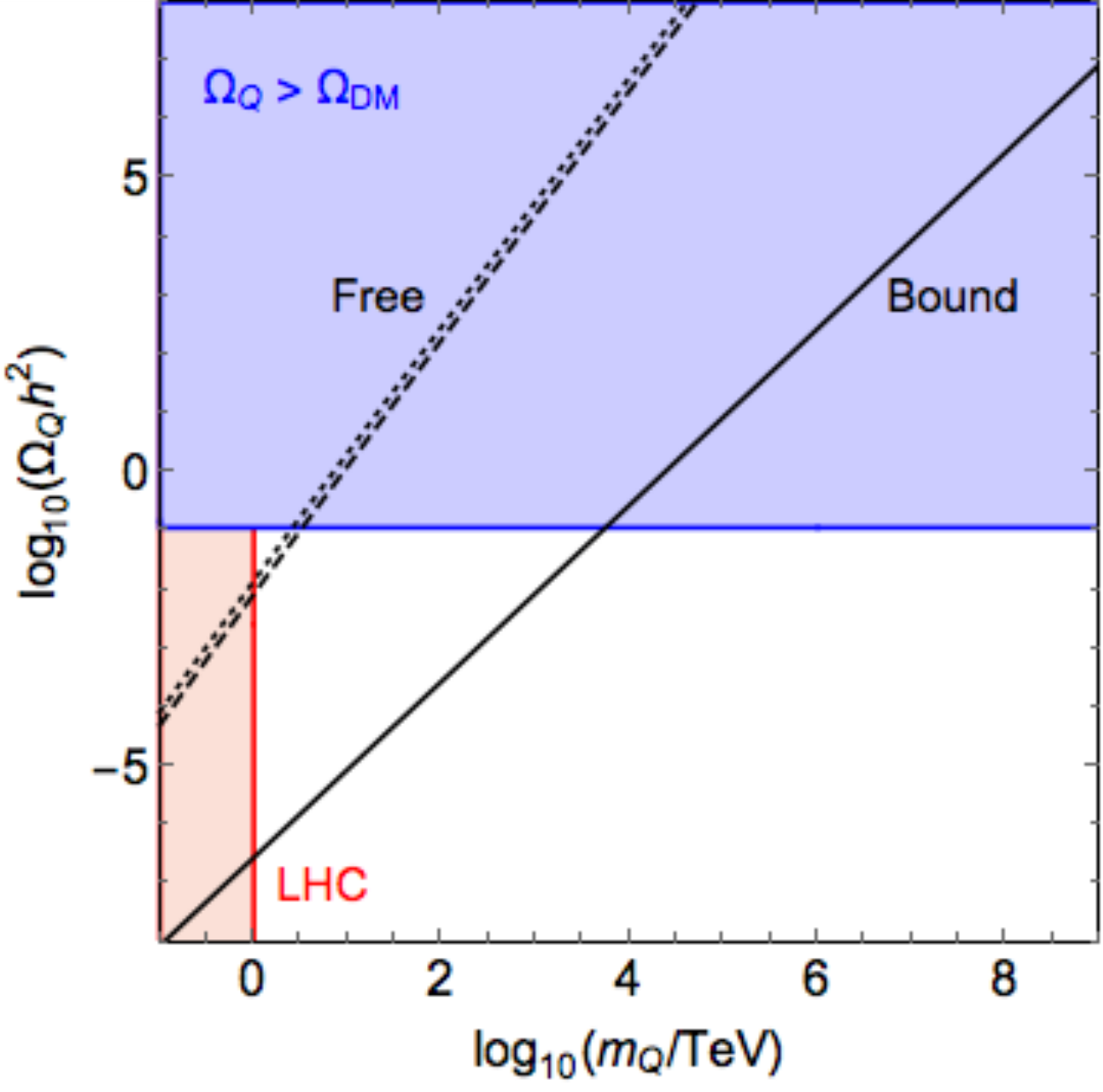}
\caption{\label{fig:overclosure} Heavy $Q$ 
contribution to the cosmological energy density versus $m_Q$.  The 
dotted (dashed) line corresponds to free annihilation for color triplets 
(octets). The solid line to annihilation via bound state formation.}
\end{figure}

\Sec{Cosmology.}  
We assume a post-inflationary scenario 
($U(1)_{PQ}$ broken after
inflation).  Then, requiring that the axion energy density from vacuum
realignment does not exceed $\Omega_{DM}$ implies
$V_a/N_{DW} \equiv f_a \lesssim f_a^{\text{max}}$, with $f_a^{\text{max}} = 5\cdot
10^{11}$ GeV~\cite{Bonati:2015vqz,Petreczky:2016vrs,Borsanyi:2016ksw},
where $N_{DW}=2N$ is the vacuum degeneracy corresponding to a
$\zb_{2N}\subset U(1)_{PQ}$ left unbroken by non-perturbative QCD
effects.
We further  assume $m_Q < T_{\rm reheating}$ so
that 
a thermal distribution of $Q$ 
provides the initial conditions for their cosmological history, which
then depends only on the mass $m_Q$ and representation $R_Q$.  For
some $R_Q$, only fractionally charged $Q$-hadrons can appear after
confinement, which also implies that decays into SM particles are
forbidden~\cite{KSVZlong}. These $Q$-hadrons must then exist today as
stable relics.  However, dedicated searches constrain the abundances
of fractionally charged particles 
relative to ordinary nucleons to
$n_Q/n_b \lsim 10^{-20}$~\cite{Perl:2009zz}, which is orders of
magnitude below any reasonable estimate of the relic abundance and of
the resulting concentrations in bulk matter. This restricts the
viable $R_Q$ to the much smaller subset for which $Q$-hadrons are
integrally charged or neutral. In this case decays into SM particles
are not forbidden, but the lifetime $\tau_Q$ is severely constrained
by cosmological observations. For $\tau_Q \sim (10^{-2}-10^{12})$\,s 
$Q$ decays would affect Big Bang Nucleosynthesis (BBN)
\cite{Kawasaki:2004qu,Jedamzik:2007qk}.  The window
$\tau_Q\sim (10^{6}-10^{12})$\,s is strongly constrained also by
limits on CMB spectral distortions from early energy
release~\cite{Hu:1993gc,Chluba:2011hw,Chluba:2013wsa}, while decays
around the recombination era ($\tau_Q\gsim 10^{13}$\,s) would leave
clear traces on CMB anisotropies. Decays after recombination would
produce free-streaming photons visible in the diffuse gamma ray
background \cite{Kribs:1996ac}, and Fermi LAT
limits~\cite{Ackermann:2012qk} allow to exclude
$\tau_Q\sim (10^{13}-10^{26})$\,s. For lifetimes longer than the age
of the Universe $\tau_Q\gsim 10^{17}$\,s the $Q$ would
contribute to the present energy density, and we must require
$\Omega_Q \leq \Omega_{DM} \approx 0.12 \, h^{-2}$.  
However, estimating  $\Omega_Q$ is not so simple. 
Before confinement the $Q$'s annihilate as free quarks.  
Perturbative calculations are reliable giving, for  
$n_f$ final state quark flavors:
\begin{equation}
  \label{eq:pertannh}
\langle \sigma v\rangle_{Q\bar Q} = 
  \frac{\pi\alpha_s^2}{16 m_Q^2}\left(c_f n_f + c_g\right) \,, 
\end{equation}
with, e.g.,  $(c_f,c_g)= (\frac{2}{9},\frac{220}{27})$ for triplets 
and $(\frac{3}{2},\frac{27}{4})$ for octets.  
Free $Q$ annihilation freezes out around $T_{fo} \sim m_Q/25$  
when (for $m_Q>$ few TeV) there are $g_* = 106.75$ 
effective degrees of freedom in thermal equilibrium. 
Together with~\eq{eq:pertannh} this gives:
\begin{equation}
\label{relicpertQCD}
\left( \Omega_{Q} h^2 \right)^{\rm Free} \approx 8 \cdot 10^{-3} 
 \left(\frac{m_Q}{{\rm TeV}}\right)^2\,. 
\end{equation}
The upper lines in \fig{fig:overclosure} give
$\left( \Omega_{Q} h^2 \right)^{\rm Free}$ as a function of $m_Q$ for
$SU(3)_C$ triplets (dotted) and octets (dashed).  Only a small corner at
low $m_Q$ satisfies $\Omega_Q\leq \Omega_{DM}$, and 
future 
improved 
LHC limits on $m_Q$ 
might exclude it completely.
However, after confinement ($T_{C} \approx 180$ MeV), because of
finite size effects of the composite $Q$-hadrons annihilation could
restart.  Some controversy exists about the possible
enhancements for annihilations in this 
regime.  For example, 
a cross section typical of inclusive hadronic scattering
$\sigma_{\rm ann} \sim (m_\pi^2 v)^{-1} \sim 30 \, v^{-1}\,$mb was
assumed in Ref.~\cite{Dover:1979sn} yielding  
$n_Q/n_b \sim 10^{-11}$.  It was later remarked~\cite{Nardi:1990ku}
that the relevant process is exclusive 
(no $Q$ quarks in the final state) with a cross section quite likely
smaller by a few orders of magnitude.  Ref.~\cite{Arvanitaki:2005fa}
suggested that bound states formed in the collision of two $Q$-hadrons
could catalyse annihilations. This mechanism was reconsidered
in~\cite{Kang:2006yd,Jacoby:2007nw} which argued that $\Omega_Q$ could
indeed be efficiently reduced.  Their results imply:
\begin{equation}
\label{relicKLNQCD}
\left( \Omega_{Q} h^2 \right)^{\rm Bound} \approx
3 \cdot 10^{-7} 
 \left( \frac{m_Q}{\text{TeV}} \right)^{3/2} \, ,
\end{equation}
which corresponds to the continuous line in \fig{fig:overclosure}.
Ref.~\cite{Kusakabe:2011hk} studied this mechanism more
quantitatively, and concluded that \eq{relicKLNQCD} represents a lower
limit on $\Omega_Q$, but much larger values are also possible. 
Refs.~\cite{Kang:2006yd,Jacoby:2007nw} in fact did not
consider the possible formation of $QQ...$ bound states which,
opposite to $Q\bar Q$, would hinder annihilation rather than catalyse
it. Then, if a sizeable fraction of $Q$'s gets bounded in such states,
the free quark result~\eqn{relicpertQCD} would give a better estimate
than~\eqn{relicKLNQCD}.  If instead the estimate~\eqn{relicKLNQCD} is
correct, energy density considerations would not exclude relics with
$m_Q\lesssim 5.4 \cdot 10^3$ TeV, nevertheless, present concentrations
of $Q$-hadrons would still be rather large $10^{-8} \lsim n_Q/n_b
\lsim 10^{-6} $.
While it has been debated if concentrations of the same order should be
expected also in the Galactic
disk~\cite{Dimopoulos:1989hk,Chuzhoy:2008zy} 
searches for anomalously
heavy isotopes in terrestrial, lunar, and meteoritic materials yield
limits on $n_Q/n_b$ many orders of magnitude below the
quoted numbers~\cite{Burdin:2014xma}.  
Moreover, even a tiny amount of heavy $Q$'s in the interior of
celestial bodies (stars, neutron stars, Earth) would produce all
sorts of effects like instabilities~\cite{Hertzberg:2016jie},
collapses~\cite{Gould:1989gw}, anomalously large heat
flows~\cite{Mack:2007xj}. 
Therefore, unless  an extremely efficient mechanism exists 
that keeps $Q$-matter completely separated from ordinary matter,
stable $Q$-hadrons would be ruled out. 

\Sec{Selection criteria.}  
The first  criterium  
 to discriminate hadronic axion models is:\ {\it $i)$ Models that allow
  for lifetimes $\tau_Q \lsim 10^{-2}\,$s are
  phenomenologically preferred with respect to models containing long
  lived or cosmologically stable $Q$'s}. All $R_Q$ allowing for decays
via renormalizable operators satisfy this requirement. Decays can also
occur via operators of higher dimensions. 
We assume that the cutoff scale is $m_P$ and write
$\mathcal{O}^{d>4}_{Qq} = m_P^{4-d} \mathcal{P}_d(Q,\varphi^n)$ where
$\mathcal{P}_d$ is a $d$-dimensional Lorentz and gauge invariant
monomial linear in $Q$ and containing $n$ SM fields $\varphi$.  For
$d=5,6,7$ the final states always contain $n\geq d-3$
particles. Taking conservatively $n=d-3$ we obtain:
\begin{equation}
\label{Gammad}
\Gamma_{d} \lsim \frac{\pi g_f m_Q}{ (d-4)!(d-5)!}  \left(\frac{m_Q^2}{16 \pi^2
    m_P^2}\right)^{d-4}\, ,   
\end{equation}
%
with $g_f$ the final degrees of freedom, 
and we have
integrated analytically the $n$-body phase space neglecting $\varphi$
masses and taking momentum independent matrix elements (see
e.g.~\cite{DiLuzio:2015oha}).  
For $d=5,6,7$ we obtain $ \tau_Q^{(d)}\gsim \left(4\cdot
  10^{-20},7\cdot 10^{-3},4\cdot 10^{15}\right) \times
(f_a^{\text{max}}/m_Q)^{2d-7}\,$s.  For $d=5$, as long as $m_Q\gsim
800$ TeV decays occur with safe lifetimes $\tau^{(5)}_Q\lsim
10^{-2}\,$s.  For $d=6$, even for the largest values $m_Q \sim
f^{\text{max}}_a$ decays occur dangerously close to
BBN~\footnote{Since $m_Q \sim y_Q N_{DW} f_a$, if $y_Q N_{DW} >1$ we
  can have $m_Q>f_a$ and a window could open up also for some $d=6$
  operators. This case will be addressed in~\cite{KSVZlong}.}.
Operators of $d=7$ and higher are always excluded.  This selects the
$R_Q$ which allow for $\mathcal{L}_{Qq}\neq 0$ (the first seven
in~\Table{summarydim5}), plus other thirteen which allow for $d=5$
decay operators.  Some of these representations are, however, rather
large, and can induce LP in the SM gauge couplings $g_1,\, g_2,\,g_3$
at some uncomfortably low-energy scale $\Lambda_{LP} < m_P$.
Gravitational corrections to the running of gauge couplings become
relevant at scales approaching $m_P$, and can delay the emergence of
LP~\cite{Robinson:2005fj}. We then specify our second criterium
choosing a value of $\Lambda_{LP}$ for which these corrections can
presumably be neglected:\ {\it $ii)$ $R_Q$'s which do not induce LP in
  $g_1,g_2,g_3$ below $\Lambda_{LP} \sim 10^{18}\,{\rm GeV}$ are
  phenomenologically preferred.}  We use two-loop beta
functions to evolve the couplings~\cite{DiLuzio:2015oha} and set
(conservatively) the threshold for $R_Q$ at $m_Q = 5 \cdot 10^{11}$
GeV. The $R_Q$ surviving this last selection are listed
in~\Table{summarydim5}.

Other features can render some $R_Q$ more appealing than others. For
example problems with cosmological domain walls~\cite{Sikivie:1982qv}
are avoided for $N_{DW}=1$, while specific $R_Q$ can
improve gauge coupling unification~\cite{Giudice:2012zp}.  We prefer
not to consider these as crucial discriminating criteria, since
solutions to the DW problem exist (see
e.g.~\cite{Kim:1986ax,Ringwald:2015dsf}), while improved unification
might be accidental because of the many $R_Q$ we consider.
Nevertheless, we have studied both these issues.  The values of
$N_{DW}$ are included in \Table{summarydim5} while, as already noted
in~\cite{Giudice:2012zp}, gauge coupling unification gets considerably
improved only for $R_3$.

\begin{table}[t!] 
\renewcommand{\arraystretch}{1.2}
\centering
\begin{tabular}{@{} |l|c|c|c|c| @{}}
\hline
$\ \ \ \ \ \ \ \ R_Q$ &  $\mathcal{O}_{Qq}$ & $\Lambda^{\!R_Q}_{LP}$[GeV] & $E/N$ & $N_{DW}$ \\ 
\hline
\hline
$R_1$:$\,(3,1,-\tfrac{1}{3})$ & 
$\overline{Q}_L d_R$ 
& $9.3 \cdot 10^{38} (g_1)$ & $2/3$ & $1$ \\ 
\hline
$R_2$:$\,(3,1,+\tfrac{2}{3})$ & 
$\overline{Q}_L u_R$
& $5.4 \cdot 10^{34} (g_1)$ & $8/3$ & $1$ \\ 
\hline
$R_3$:$\,(3,2,+\tfrac{1}{6})$ & 
$\overline{Q}_R q_L$
& $6.5 \cdot 10^{39} (g_1)$ & $5/3$ & $2$ \\ 
\hline
$R_4$:$\,(3,2,-\tfrac{5}{6})$ & 
$\overline{Q}_L d_R H^\dag$
& $4.3 \cdot 10^{27} (g_1)$ & $17/3$ & $2$ \\
\hline
$R_5$:$\,(3,2,+\tfrac{7}{6})$ & 
$\overline{Q}_L u_R H$
& $5.6 \cdot 10^{22} (g_1)$ & $29/3$ & $2$ \\
\hline
$R_6$:$\,(3,3,-\tfrac{1}{3})$ & 
$\overline{Q}_R q_L H^\dag$ 
& $5.1 \cdot 10^{30} (g_2)$ & $14/3$ & $3$ \\
\hline
$R_7$:$\,(3,3,+\tfrac{2}{3})$ & 
$\overline{Q}_R q_L H$ 
& $6.6 \cdot 10^{27} (g_2)$ & $20/3$ & $3$ \\
\hlineB{2.5}
$R_8$:$\,(3,3,-\tfrac{4}{3})$ & 
$\overline{Q}_L d_R H^{\dag 2}$ & $3.5 \cdot 10^{18} (g_1)$ & $44/3$ & $3$ \\
\hline
$R_9$:$\,(\bar 6,1,-\tfrac{1}{3})$ & 
$\overline{Q}_L \sigma  d_R \cdot G $ & $2.3 \cdot 10^{37} (g_1)$ & $4/15$ & $5$ \\
\hline
$R_{10}$:$\,(\bar 6,1,+\tfrac{2}{3})$ & 
$\overline{Q}_L \sigma  u_R \cdot G $ & $5.1 \cdot 10^{30} (g_1)$ & $16/15$ & $5$ \\
\hline
$R_{11}$:$\,(\bar 6,2,+\tfrac{1}{6})$ & 
$\overline{Q}_R \sigma  q_L \cdot G $ & $7.3 \cdot 10^{38} (g_1)$ & $2/3$ & $10$ \\
\hline
$R_{12}$:$\,(8,1,-1)$ & 
$\overline{Q}_L \sigma  e_R \cdot G $ & $7.6 \cdot 10^{22} (g_1)$ & $8/3$ & $6$ \\
\hline
$R_{13}$:$\,(8,2,-\tfrac{1}{2})$ & 
$\overline{Q}_R \sigma  \ell_L \cdot G $ & $6.7 \cdot 10^{27} (g_1)$ & $4/3$ & $12$ \\
\hline
$R_{14}$:$\,(15,1,-\tfrac{1}{3})$ & 
$\overline{Q}_L \sigma  d_R \cdot G $ & $8.3 \cdot 10^{21} (g_3)$ & $1/6$ & $20$ \\
\hline
$R_{15}$:$\,(15,1,+\tfrac{2}{3})$ & 
$\overline{Q}_L \sigma  u_R \cdot G $ & $7.6 \cdot 10^{21} (g_3)$ & $2/3$ & $20$ \\
\hline
  \end{tabular}
  \caption{\label{summarydim5} 
    $R_Q$  allowing for $d\leq 4$ and $d=5$ decay operators 
    ($\sigma\cdot G \equiv \sigma_{\mu\nu} G^{\mu\nu}$) and yielding LP above $10^{18}$GeV.
The anomaly contribution to $g_{a\gamma\gamma}$ is given in the fourth
column, and the DW number in the fifth one.} 
\end{table}

\begin{figure}[t!]
\includegraphics[width=.48\textwidth]{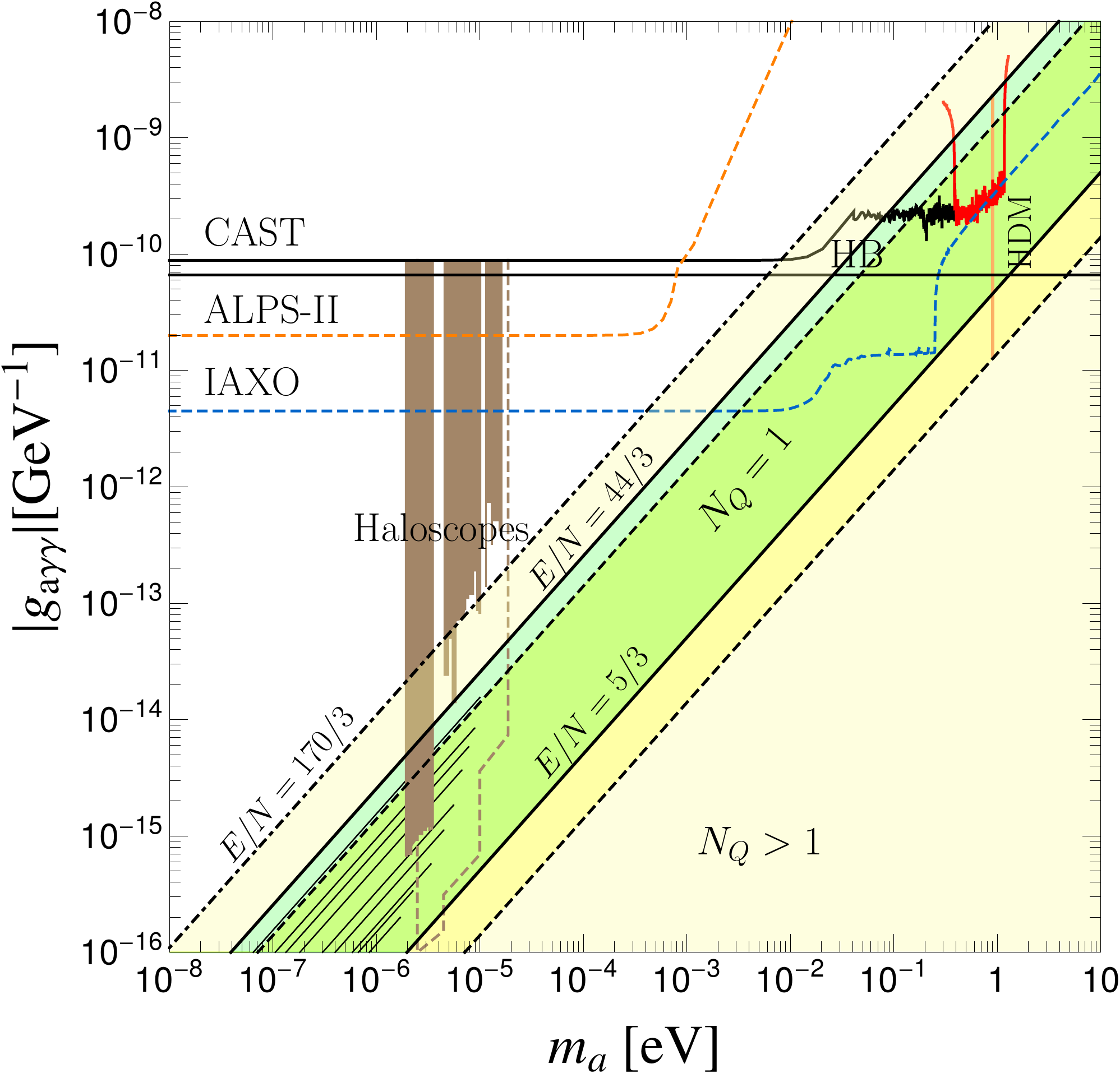}
\caption{\label{fig:Excl}
The window for preferred axion models.
The green band  encompasses models with a single $R_Q$. 
With more $R_Q$'s  the region below the  line $E/N=170/3$ 
becomes allowed.  The two dashed lines enclose     
the usual window $|E/N-1.92|\in [0.07,7]$~\cite{Olive:2016xmw}.
Current (projected) exclusion limits 
are delimited  by  solid (dashed) lines.  
}
\end{figure}

\Sec{Axion coupling to photons.} 
The most promising way to unveil the axion is via its
interaction with photons 
$ g_{a\gamma\gamma}\, a\, \mathbf{E}\cdot \mathbf{B} $ where~\cite{Kaplan:1985dv}:  
\begin{equation} 
 \label{aphcoupapprox}
g_{a\gamma\gamma}= \frac{m_a}{\rm{eV}} \ 
 \frac{2.0}{10^ {10} \rm{GeV}} \; \left(\frac{E}{N} - 1.92(4)\right)\,,  
\end{equation}
with $N,\,E$ the anomaly coefficients in~\eqssmall{N}{E} (the uncertainty
comes from the NLO chiral Lagrangian~\cite{diCortona:2015ldu}). The last
column in \Table{summarydim5} gives $E/N$ for the selected $R_Q$'s.
We have sketched in~\fig{fig:Excl} the ``density'' of preferred hadronic
axion models, drawing with oblique lines (only at small $m_a$)
the corresponding couplings.  The $s$trongest coupling is obtained for
$R^s_Q=R_8$ and the $w$eakest for $R^w_Q=R_3$. They  
delimit a  window $0.25 \leq |E/N-1.92|\leq 12.75$ encompassing all
axion models in~\Table{summarydim5}. 
The corresponding couplings $g_{a\gamma\gamma}$ fall within the band delimited
in \fig{fig:Excl} by the lines $E/N=5/3$ and $44/3$.  
With respect to the usual window 
 $0.07 \leq |E/N-1.92|\leq 7$~\cite{Olive:2016xmw}
(delimited 
by the two 
dashed lines) the upper (lower) limit is shifted upwards
approximatively by a factor of 2 (3.5).
It is natural to ask if $g_{a\gamma\gamma}$ could get enhanced by
allowing for more $R_Q$'s  ($N_Q>1$). Let us consider the combined anomaly
factor for $R^s_Q \oplus R_Q$:
\beq 
\label{EcNc}
\frac{E_c}{N_c}\equiv 
\frac{E+E_s}{N+N_s} = \frac{E_s}{N_s} \left(\frac{1 +
    E/E_s}{1+N/N_s}\right) \, . 
\eeq 
Since by construction the anomaly coefficients of all $R_Q$'s in our
set satisfy $E/N \leq E_s/N_s $, the factor in parenthesis is $\leq 1$
implying $E_c/N_c \leq E_s/N_s$. This result is easily generalized to
$N_Q>2$. Therefore, as long as the sign of $\Delta\mX = \mX_L-\mX_R$
is the same for all $R_Q$'s, no enhancement is possible.  However, if
we allow for $R_Q$'s with PQ charge differences of opposite sign (we
use the symbol $\ominus$ to denote reducible representations of this
type)
$E/E_s$ and $N/N_s$ in \eq{EcNc} become negative and
$g_{a\gamma\gamma}$ can get enhanced.  For $N_Q=2$ the largest value
is $E_c/N_c = 122/3$ obtained for $R_Q^s \ominus R_Q^w$.  For $N_Q>2$
even larger couplings can be obtained. However, contributions to the
$\beta$-functions also become large and can induce LP.  This implies
that there is a maximum value $g^{\rm max}_{a\gamma\gamma}$ for which
our second condition remains satisfied.  We find that $R^s_Q\oplus
R_6\ominus R_9$, giving $E_c/N_c= 170/3$, yields the largest possible
coupling. The uppermost oblique line in~\fig{fig:Excl} depicts 
the corresponding $g^{\rm max}_{a\gamma\gamma}$.  More $R_Q$'s can also
suppress $g_{a\gamma\gamma}$ and even produce a complete decoupling.
This requires an ad hoc choice of $R_Q$'s, but no numerical fine
tuning.  With two $R_Q$'s there are three cases yielding
$g_{a\gamma\gamma}=0$ within theoretical errors~\cite{KSVZlong}
(e.g.~$R_6 \oplus R_9$ giving $E_c/N_c = 23/12\simeq 1.92$).  This
provides additional motivations for search techniques which do not
rely on the axion coupling to photons~\cite{Budker:2013hfa,Arvanitaki:2014dfa}.
Finally, since $T(8)=3$ and  
$T(6)=5/2$,  by combining with opposite PQ charge differences 
$R_{12}$ with $R_9$ or $R_{10}$,  
new models with $N_{DW}=1$  can be constructed.


We have classified hadronic axion models using  
well-defined phenomenological criteria. 
The window of preferred models is shown in 
\fig{fig:Excl}. 

\section*{Acknowledgments}

We thank D.~Aristizabal Sierra for several discussions  
since the early stages of this work, and   
M.~Giannotti, S.~Nussinov and  J.~Redondo  
for useful feedbacks. 
F.M.~acknowledges financial support from FPA2013-46570, 2014-SGR-104 and MDM-2014-0369.  
E.N.~is supported by the research grant
No.~2012CPPYP7 of the MIUR program PRIN-2012, 
and by the INFN ``Iniziativa Specifica'' TAsP.

 \bibliographystyle{apsrev4-1.bst}
 \bibliography{bibliography}

\end{document}